%% file: aa_2005_4405.tex
\documentclass{aa}
\usepackage{times}
\usepackage{graphicx}
\usepackage{xspace}
\usepackage{epsfig}
\usepackage{natbib}
\usepackage{rotating}

\begin{document}

\title{Testing the companion hypothesis for the origin of the \\ X-ray emission from intermediate-mass main-sequence stars}

\author{B. Stelzer\inst {1} \and N. Hu\'elamo\inst {2} \and G. Micela\inst {1} \and S. Hubrig\inst {2}} 

\offprints{B. Stelzer}

\institute{INAF - Osservatorio Astronomico di Palermo,
  Piazza del Parlamento 1, I-90134 Palermo, Italy \\ \email{B. Stelzer, stelzer@astropa.unipa.it}
\and
  European Southern Observatory, 
  Casilla 19001, Santiago 19, 
  Chile} 

\titlerunning{X-ray emission from main-sequence B-stars}

\date{Received $<$25-10-2005$>$ / Accepted $<$20-02-2006$>$}

\abstract
{The X-ray emission from B-type main-sequence stars is a longstanding
mystery in stellar coronal research. Since there is no theory at hand that explains
intrinsic X-ray emission from intermediate-mass main-sequence stars, the observations
have often been interpreted in terms of (unknown) late-type magnetically active companion stars.}
{Resolving the hypothesized companions requires high spatial resolution observations
in the infrared and in X-rays. We use Chandra imaging observations to spatially resolve a sample of 
main-sequence B-type stars with recently discovered companions at arcsecond separation.
} 
{Our strategy is to search for X-ray emission at the position of both the B-type primary and the faint 
companion. 
}
{We find that all spatially resolved companions are X-ray emitters, 
but seven out of eleven intermediate-mass stars are also X-ray sources. 
If this emission is interpreted in terms
of additional sub-arcsecond or spectroscopic companions, this implies a high multiplicity
of B-type stars. Firm results on B star
multiplicity pending, the alternative, that B stars produce intrinsic X-rays, can not be discarded. 
The appropriate scenario in this vein is probably a magnetically confined wind, as suggested 
for the X-ray emission of the magnetic Ap star IQ\,Aur. 
However, the only Ap star in the Chandra sample is not detected in X-rays,
and therefore does not support this picture.}
{}


\maketitle

\section{Introduction}\label{sect:intro}

After their discovery with the {\em Einstein} satellite \citep{Cash82.1, Caillault89.1} 
X-ray detections of A- and B-type stars have repeatedly been reported
throughout the literature \citep[see references in ][]{Stelzer03.1},
but their origin has remained a mystery. 

Intermediate-mass stars are not expected to give rise to high-energy
emission. In contrast to early-type stars they do not drive strong
stellar winds, in which instabilities arise that can be 
held responsible for the X-rays such as in O- and early B-type stars 
\citep{Owocki99.1, Lucy80.1}. The empirical relation between
the bolometric and the X-ray luminosity observed for hot stars breaks down near 
spectral type $\sim$B2 \citep[][ hereafter BSC96]{Berghoefer96.1}, approximately coincident
with the transition from a strong-wind to a weak-wind regime \citep{Babel96.1}.
Stars with cooler spectral types show 
a large scatter of X-ray luminosity for given bolometric luminosity. 

Intermediate-mass stars also do not possess convection
zones necessary to drive a stellar dynamo. Therefore, magnetic activity
which generates X-ray emission of late-type stars 
\citep{Parker55.1, Parker93.1, Ruediger95.1}
seems not to be a viable alternative. 
The minimum depth of a convective envelope able to support magnetic 
activity is not well established. 
Based on studies of the X-ray detection rate as a function of color for
main-sequence (MS) stars the onset of significant dynamo
action is placed somewhere near spectral type A7 
\citep{Schmitt85.1, Huensch01.1}. 

As outlined above, obvious qualitative changes in the properties of the
observed X-ray emission coincide with the approximate break-down of
known emission mechanisms. Nevertheless, 
contrary to the expectation a relatively large number of A- and B-type stars apparently are
not X-ray dark, with hundreds of them listed in the catalogs 
of X-ray sources detected during the {\em ROSAT} All-Sky Survey (RASS);  
see \citet{Berghoefer96.1, Huensch98.1}. 
In absence of a physical explanation, the X-ray emission of late-B 
and early-A stars is commonly attributed to unresolved late-type companions.
The contraction timescale of late-type pre-MS stars to the MS is comparable to the 
mean lifetime of late B-type stars on the MS. Hence, the hypothesized companions 
of a MS B-type star must be young late-type stars in approach to the MS, and
-- because of the relation between X-ray emission and stellar youth -- 
they are by their nature strong X-ray sources. 
Based on this argument \citet{Lindroos86.1} has isolated $78$ pairs 
composed of a MS early-type star and a likely physically bound pre-MS star. 
These systems are henceforth termed `Lindroos systems'. 

Testing the hypothesis that unresolved companions produce the observed X-rays from
A- and B-type stars requires high-spatial resolution observations in the optical/IR
and the X-ray band. The first step in this approach is to search for companions 
within the X-ray error box of X-ray detected A/B-stars. The preferred wavelength is the 
near-IR, where the contrast between the intermediate-mass star and an eventual cool 
companion is more favorable with respect to the optical, and where the adaptive optics (AO) 
technique allows to obtain diffraction limited images with sub-arcsecond separations. 
This way \citet{Hubrig01.1} reported the discovery of new
companions to $19$ of $49$ B-type stars selected among the X-ray emitters detected during
the RASS. The separations between the new IR objects and 
the B-type primary range between $0.2-14^{\prime\prime}$.
The IR photometry of these new objects has been used to place them in the HR
diagram, estimating their masses and spectral types. Both are
consistent with them being late-type stars. Therefore, they are good candidates 
for the origin of the X-ray emission. 

After the identification of the companions the claim that they are responsible for
the observed X-ray emission must be proven by resolving them in X-rays from the 
intermediate-mass stars. So far, observational investigations of this issue have been
restricted to the few known A- and B-type stars with companions at
separations large enough to be resolved by the low spatial resolution
X-ray instruments available.
A sample of eight visual binaries with separations $>\,10^{\prime\prime}$ was examined
by \citet{Berghoefer94.1} using the {\em ROSAT} High Resolution Imager (HRI). 
But in only one case the X-ray emission could be ascribed to the known visual
late-type companion. On the other hand, a {\em ROSAT} HRI study of the Lindroos
systems has shown that both the primaries -- mostly of spectral type B -- and their
late-type companions emit X-rays at similar levels 
\citep{Schmitt93.2, Huelamo00.1}. This was taken as support for the 
hypothesis that the X-ray emission from the late B-type stars in fact 
originates from closer pre-MS late-type companions unresolvable by the
{\em ROSAT} HRI.

{\em Chandra} is the first X-ray satellite that allows for a spatial
resolution which comes near to that in the IR. Systems as close as 
$\sim 1^{\prime\prime}$ can be studied. At the distance of $100-200$\,pc where many 
of the X-ray emitting B- and A-type stars are located, this corresponds to 
$< 200$\,AU separation. This is much smaller than the maximum separation of
visual binaries in the solar neighborhood 
\citep[see ][]{Close90.1, Duquennoy91.1}. 
Therefore, it is plausible to consider faint objects found at 
a separation of few arc-seconds from nearby B- and A-type stars bound companions, and not just
 chance projections. However, for the main purpose of this study (i.e. the investigation
of the origin of the X-rays from B-type stars) this question is not of great importance.  

We have performed a series of {\em Chandra} imaging observations pointing at 
multiple stars with a B-type primary selected from recent AO surveys. 
The present article is a continuation of the work presented by \citet{Stelzer03.1}
(henceforth SHH03). 
In Sect.~\ref{sect:sample} we explain the selection of the sample
observed with {\em Chandra}. In Sect.~\ref{sect:observations} the observations and 
the data analysis are described.
The results from source detection and the X-ray properties of the
detections are also presented in Sect.~\ref{sect:observations},
including a discussion of the X-ray lightcurves and spectra. 
We discuss the $L_{\rm x}/L_{\rm bol}$ relation in Sect~\ref{sect:lxlbol}. 
In Sect.~\ref{sect:discussion} the results are put in the context of
possible emission mechanisms, and the appendix~\ref{sect:newdata} has
information on the individual targets.

\section{The Sample}\label{sect:sample}

Our project consists of {\em Chandra} snapshots of 
late-B type stars detected in the RASS and known to have close faint companions
revealed in recent near-IR AO observations. 
Results for the first 5 objects were presented by SHH03, 
where our selection criteria have been described in detail. 
Briefly summarized, the targets must (i) have been detected previously in
low-spatial resolution X-ray observations, (ii) have companions identified by recent 
AO observations and unknown at the time of the previous X-ray observations, 
(iii) have separations ranging from $1-8^{\prime\prime}$ between the B/A star and the
AO companion  
(to be resolvable with {\em Chandra} but no other
presently or previously available X-ray instrument), 
and (iv) the primaries show no signs of intrinsic binarity according to the
{\em Hipparcos} data base (The Hipparcos Catalogue, 1997, ESA SP-1200) 
and the $\Delta\mu$ data base \citep{Wielen00.1}.

In this paper the sample presented by SHH03 is being complemented by {\em Chandra} observations
of another $5$ late-B stars.  
Our targets were chosen from the work of \citet{Hubrig01.1}.
We have added to our sample one further intermediate-mass star from the {\em Chandra} 
archive: HD\,129791 is a Lindroos system composed of a A0 star on the MS and
a K5 companion at $35^{\prime\prime}$. 
One of our other targets, HD\,32964, is also a Lindroos system,
i.e. this star has both a wide companion at $\sim 53^{\prime\prime}$ (the Lindroos secondary) 
and a close companion discovered by \citet{Hubrig01.1}. 

The separations of the two Lindroos secondaries are 
large enough so that they have been resolved with {\em ROSAT}. Therefore, these companions 
can not provide the explanation of the X-ray emission from the B-type primaries. 
However, studying their X-ray emission may be instructive for two reasons: 
(1) It may help to discriminate between them being physically bound 
companions and unbound chance projections. This is, because -- as explained in Sect.~\ref{sect:intro} -- 
true late-type companions to B stars on the MS are still in their pre-MS contraction phase
where they are strongly magnetically active, while unrelated old foreground or background 
stars are not expected to be X-ray bright. (2) It allows to search for 
differences between primaries and secondaries that could point at different emission
processes.  

To summarize, the new sample we present in this paper comprises $6$ targets, $5$ from
our dedicated {\em Chandra} program and $1$ from the {\em Chandra} archive. 
In Table~\ref{tab:obslog} some information on the targets is summarized.  
Cols.~$1-5$ give the {\em Hipparcos} position of the primary, distance, and spectral type. 
The separation, position angle, an identifier, and the reference  
of all known companions are listed in cols.~$6-9$. We use labels `L' for Lindroos companions,
and letters `B' and `C' for further visual companions.  
The last columns of Table~\ref{tab:obslog} comprise the {\em Chandra} 
observation identification number, instrument, and exposure time.  
Throughout this paper we also use results from the first part of our survey (SHH03).
The targets presented by SHH03 are also listed in Table~\ref{tab:obslog}.
Where necessary, the X-ray properties of that group of stars have been re-examined
to yield a consistent description of the whole sample.  

We point out that true companionship has not been established for any of
the new IR objects in our sample, and only for some of the Lindroos secondaries. 
Confirmation that they are true companions and not just chance projections 
requires observations of their proper motion and/or spectra. 
In absence of definite information about their status
we will for simplicity continue calling the IR objects `companions', and the 
B-type stars `primaries'.

%
%
\begin{table*}\small
\begin{center}
\caption{Target list of {\em Chandra} observed late-B and early-A stars: position, distance, and spectral type; separation, position angle, and an identifier of companion candidates; {\em Chandra} observation ID, detector ID, and exposure time.}
\label{tab:obslog}
\begin{tabular}{lrrrlrrccrcr}
\noalign{\smallskip} \hline \noalign{\smallskip} 
Designation & \multicolumn{2}{c}{Position$^{(a)}$}                                       & Dist$^{(a)}$ & SpT$^{(b)}$ & \multicolumn{1}{c}{Sep$^{(c)}$}          & \multicolumn{1}{c}{PA$^{(c)}$} & Comp. & Ref. & \multicolumn{3}{c}{ACIS observations} \\
            & \multicolumn{1}{c}{$\alpha_{2000}$} & \multicolumn{1}{c}{$\delta_{2000}$}  & [pc]         &             & \multicolumn{1}{c}{[$^{\prime\prime}$]}  & \multicolumn{1}{c}{[$^\circ$]} &       &      & ObsID & Instr & Expo [s] \\
\noalign{\smallskip} \hline \noalign{\smallskip} 
\multicolumn{12}{c}{New Targets} \\
\noalign{\smallskip} \hline \noalign{\smallskip} 
HD\,32964       & 05:06:45.67 & $-$04:39:17.4 & $ 86$     & B9         & $1.61$ & $307.1$ & B & (1*) & 3739 & ACIS-I & $2518$ \\
                &             &               &           &              & $52.8$ & $10$    & L & (2)  &      &        &        \\
\noalign{\smallskip} \hline \noalign{\smallskip}
HD\,73952       & 08:38:44.96 & $-$53:05:26.1 & $155$     & B8           & $1.16$ & $205.3$ & B & (1)  & 3740 & ACIS-I & $7630$ \\
\noalign{\smallskip} \hline \noalign{\smallskip} 		     
HD\,110073      & 12:39:52.56 & $-$39:59:14.8 & $109$     & B8           & $1.19$ & $ 75.0$ & B & (1)  & 3741 & ACIS-I & $3241$ \\
\noalign{\smallskip} \hline \noalign{\smallskip} 		     
HD\,129791      & 14:45:57.66 & $-$44:52:02.9 & $130$     & A0           & $35.3$ & $205.5$ & L & (2)  & 0627 & ACIS-S & $6584$ \\
\noalign{\smallskip} \hline \noalign{\smallskip} 
HD\,134837      & 15:13:07.69 & $-$36:05:28.8 & $111$     & B8           & $4.70$ & $154.3$ & B & (1)  & 3742 & ACIS-I & $2929$ \\
\noalign{\smallskip} \hline \noalign{\smallskip} 		     
HD\,134946      & 15:13:17.44 & $-$24:00:30.2 & $126$     & B8           & $8.21$ & $ 45.3$ & B & (1)  & 3743 & ACIS-I & $2335$ \\
\noalign{\smallskip} \hline \noalign{\smallskip} 
\multicolumn{12}{c}{Stars from Stelzer et al. (2003)} \\
\noalign{\smallskip} \hline \noalign{\smallskip} 
HD\,1685        & 00:20:39.03 & $-$69:37:29.7 & $ 94$     & B9           & $2.28$ & $211.4$ & B & (1)  & 2541 & ACIS-I & $2338$ \\
\noalign{\smallskip} \hline \noalign{\smallskip} 		     
HD\,113703      & 13:06:16.70 & $-$48:27:47.8 & $127$     & B5           & $1.55$ & $268.2$ & B & (3)  & 0626 & ACIS-S &$12184$ \\
                &             &               &           &              & $11.5$ & $   79$ & L & (4) & & & \\
\noalign{\smallskip} \hline \noalign{\smallskip} 		     
HD\,123445      & 14:08:51.89 & $-43$:28:14.8 & $218$     & B9           & $5.56$ & $ 65.0$ & B & (5)  & 2542 & ACIS-I & $2237$ \\
                &             &               &           &              & $5.38$ & $ 64.0$ & C & (5)  &      &        &        \\
                &             &               &           &              & $28.6$ & $ 35$   & L & (2)  &      &        &        \\
\noalign{\smallskip} \hline \noalign{\smallskip} 		     
HD\,133880      & 15:08:12.12 & $-$40:35:02.1 & $126$     & B8           & $1.22$ & $109.2$ & B & (1)  & 2543 & ACIS-I & $2461$ \\
\noalign{\smallskip} \hline \noalign{\smallskip} 		     
HD\,169978      & 18:31:22.43 & $-$62:16:41.9 & $147$     & B7.5         & $3.09$ & $168.7$ & B & (1)  & 2544 & ACIS-I & $2420$ \\
\noalign{\smallskip} \hline \noalign{\smallskip}
\multicolumn{12}{l}{$^{(a)}$ {\em Hipparcos} position and distance for the A/B-type star} \\
\multicolumn{12}{l}{$^{(b)}$ spectral types adopted from the SIMBAD database at http://simbad.u-strasbg.fr/Simbad;} \\
\multicolumn{12}{l}{$^{(c)}$ separations have been measured in the detector space;} \\ 
\multicolumn{12}{l}{References: (1) - \protect\citet{Hubrig01.1}, (1*) - P.A. redetermined on the ADONIS image, (2) - \protect\citet{Turon93.1},} \\
\multicolumn{12}{l}{(3) - \protect\citet{Shatsky02.1}, (4) - Hu\'elamo et al, in prep., (5) - \protect\citet{Huelamo01.1}.} \\
\end{tabular}
\end{center}
\end{table*}

\section{Observations and Data Analysis}\label{sect:observations}

Our original targets were observed 
with the standard imaging array of the Advanced CCD Imaging Spectrometer 
(ACIS-I) with nominal frame time of $3.2$\,sec.  
The object added from the archive, HD\,129791, had been placed on
an ACIS-S chip. ACIS-S provides higher sensitivity than ACIS-I for soft
energies. However, its 
limiting optical magnitude is fainter. Therefore, in order to avoid the risk of 
contamination by optical light from the optically bright primary star the 
frame time had been reduced to $0.9$\,s. 
The net exposure time per target was $\sim 2 - 8$\,ks. 

The data analysis was carried out using the CIAO software 
package\footnote{CIAO is made available by the CXC and can be downloaded 
from \\ http://asc.harvard.edu/ciao/download} version 3.2
in combination with the calibration database (CALDB) version 3.0.0.
We started our analysis with the level\,1 events file provided by the
pipeline processing at the {\em Chandra} X-ray Center (CXC). 
On all observations processed at the CXC with CALDB version earlier than 2.9 
we applied a new gain map and updates 
on the geometry (focal length, ACIS pixel size and chip positions). For 
observations processed with a later version of CALDB 
these modifications had been performed automatically during the pipeline process.  
In the process of converting the level\,1 events file to a level\,2 events file
for each of the observations we performed the following steps: 
A correction for the charge transfer inefficiency (CTI) has been applied for data
with processing version earlier than 6.12, i.e. for those where the CTI correction
was not yet part of the standard pipeline processing at the CXC. 
We removed the pixel randomization which is automatically applied by the CXC pipeline
in order to optimize the spatial resolution. 
We filtered the events file for event grades
(retaining the standard {\em ASCA} grades $0$, $2$, $3$, $4$, and $6$), 
and applied the standard good time interval file. 
Events flagged as cosmic ray afterglow 
were retained after inspection of the images revealed that a substantial
number of source photons erroneously carry this flag. 
Since the positional accuracy is particularly
important to our observations we also checked the astrometry for any known 
systematic aspect offsets using the CIAO aspect calculator\footnote{see http://asc.harvard.edu/cal/ASPECT/fix\_offset/fix\_offset.cgi}. 
Small offsets are present in some of the observations, and  
the aspect was corrected accordingly by modifying the respective header keywords 
in the events level\,2 file.

\subsection{Source Detection and Identification}\label{subsect:srcdet_and_iden}

Source detection was carried out with the {\sl wavdetect} algorithm \citep{Freeman02.1}.
This algorithm correlates the data with a mexican hat function
to search for deviations from the background. The {\sl wavdetect} 
mechanism is well suited for separating closely spaced point sources.  
We used wavelet scales between $1$ and $8$ in steps of $\sqrt{2}$. 
The size of the analysed images is $50 \times 50$ sky pixels (1\,sky pixel $= 0.492^{\prime\prime}$).
For exceptions with known companions at wider separations the source
detection was performed in a slightly larger area. 
The threshold for the significance of the detection was set to $2\,10^{-7}$.
For this value the detection of one spurious source is expected in a 
$2048 \times 2048$ pixel wide image. 

The identification of all X-ray sources with components of our target systems is
summarized in cols.~$1-3$ of Table~\ref{tab:xrayparams_lx}. 
We defer a discussion of Table~\ref{tab:xrayparams_lx} to Sect.~\ref{subsect:xray_prop}.
Source identification was done by measuring the distance ($\Delta_{\rm ox}$) between each
X-ray source and the optical/IR positions of all known visual components in the system.
Then we assigned each X-ray source to the closest of the optical/IR objects. This
procedure turned out to be unambiguous for all spatially resolved components.
No limit was put on the maximum acceptable $\Delta_{\rm ox}$, but in practice these
offsets are small (typically within $0.2^{\prime\prime}$).  
As a rule, source photons were extracted from a circle centered on the
{\sl wavdetect} source position within the radius that includes 
$90$\,\% of the point-spread-function (PSF) at $1.5$\,keV, depending on the position of 
the source on the detector. 
For sources which are not fully resolved we defined smaller
non-overlapping photon extraction circles.

\subsection{Results from Source Detection}\label{subsect:results}

\begin{figure*}
\begin{center}
\parbox{18cm}{
\parbox{6cm}{
\resizebox{6cm}{!}{\includegraphics{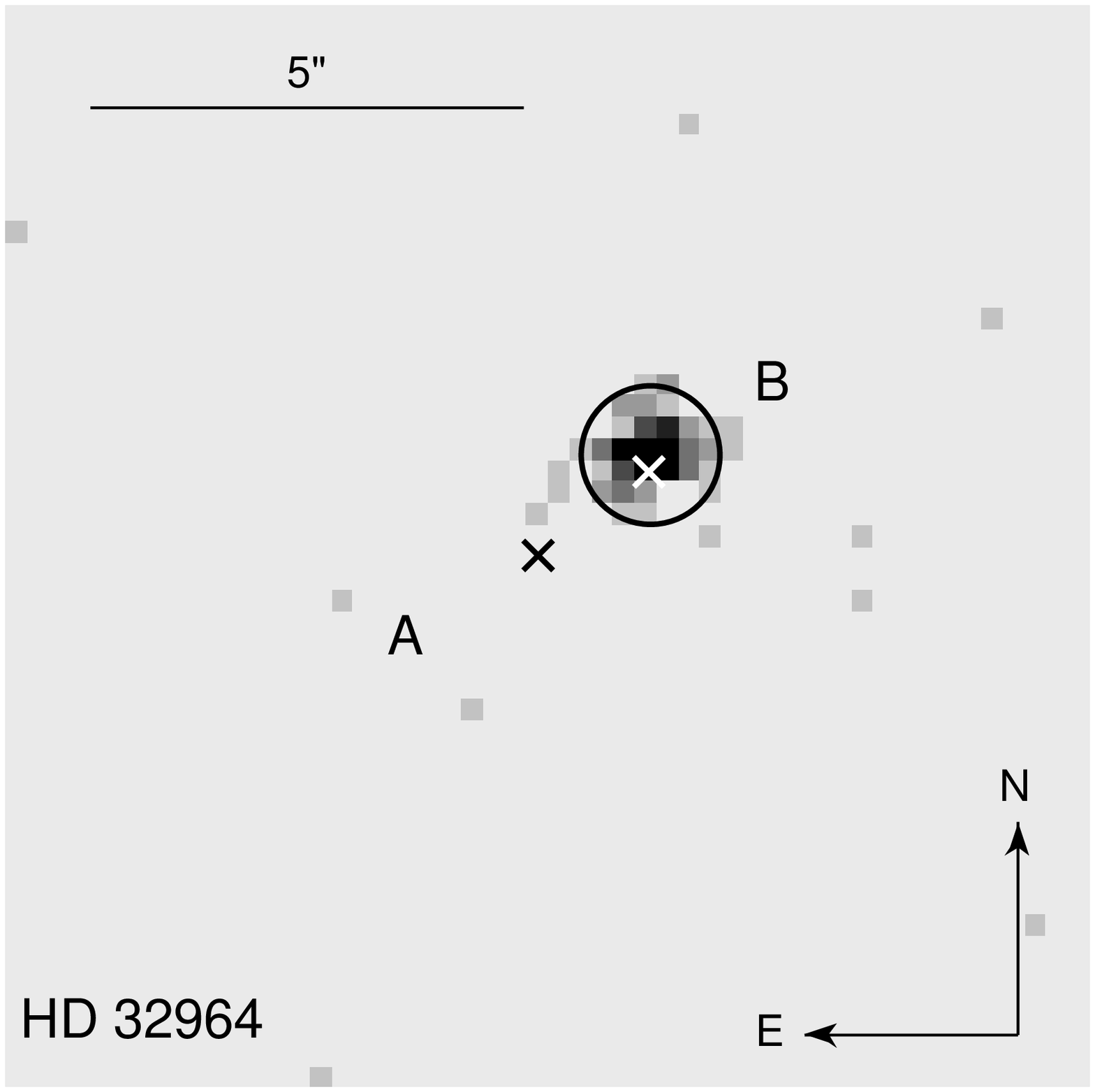}}
}
\parbox{6cm}{
\resizebox{6cm}{!}{\includegraphics{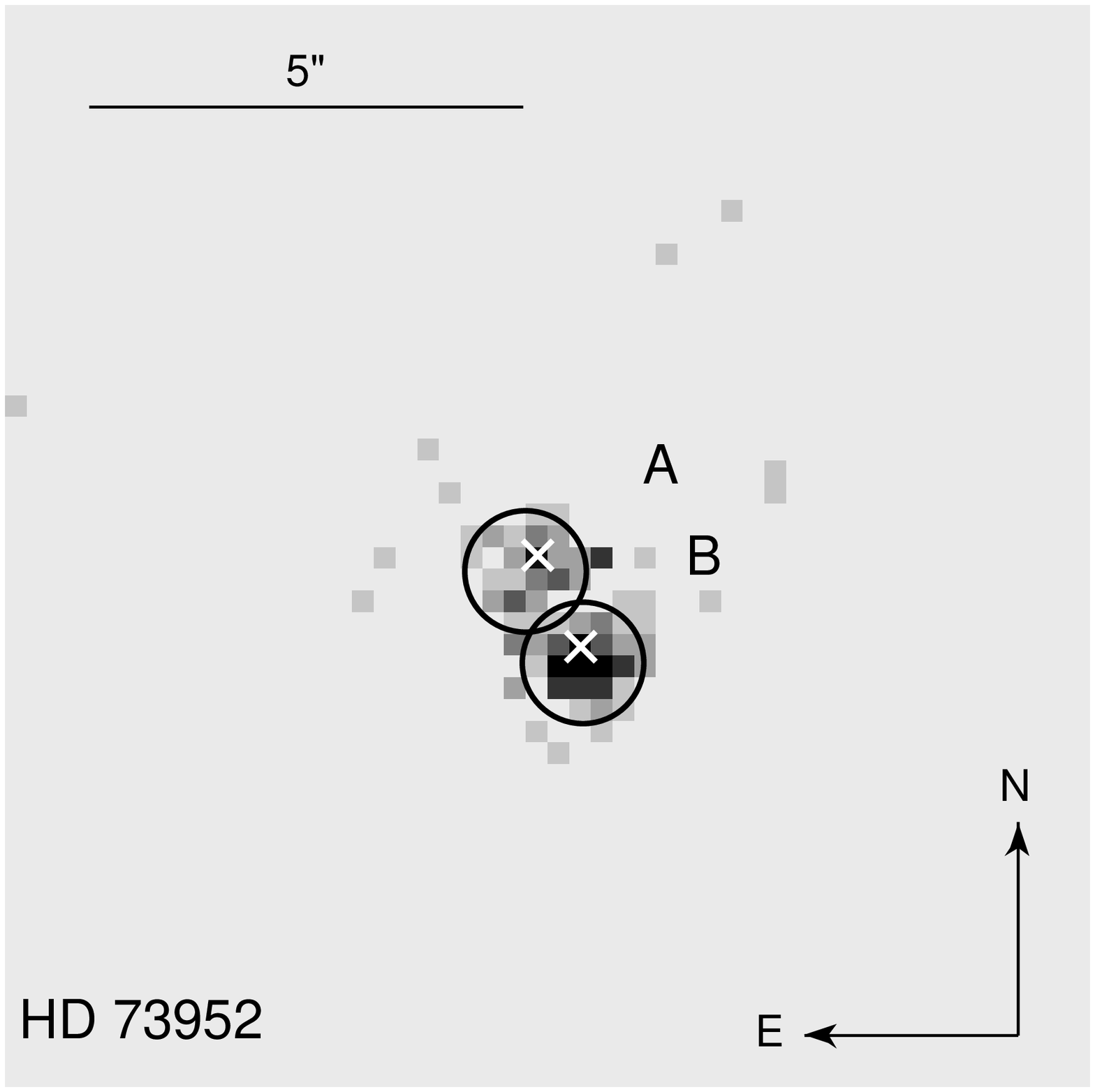}}
}
\parbox{6cm}{
\resizebox{6cm}{!}{\includegraphics{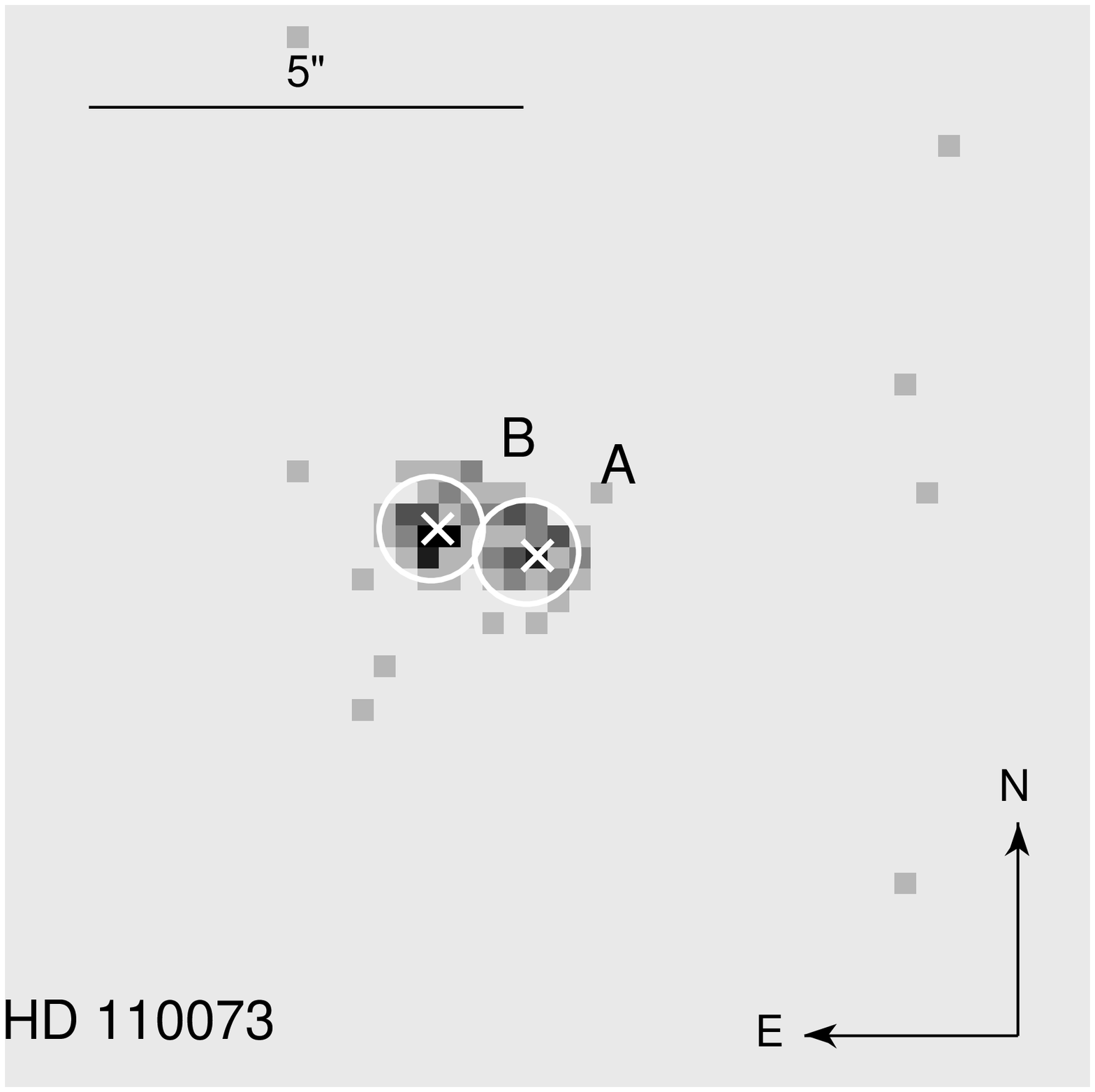}}
}
}
\parbox{18cm}{
\parbox{6cm}{
\resizebox{6cm}{!}{\includegraphics{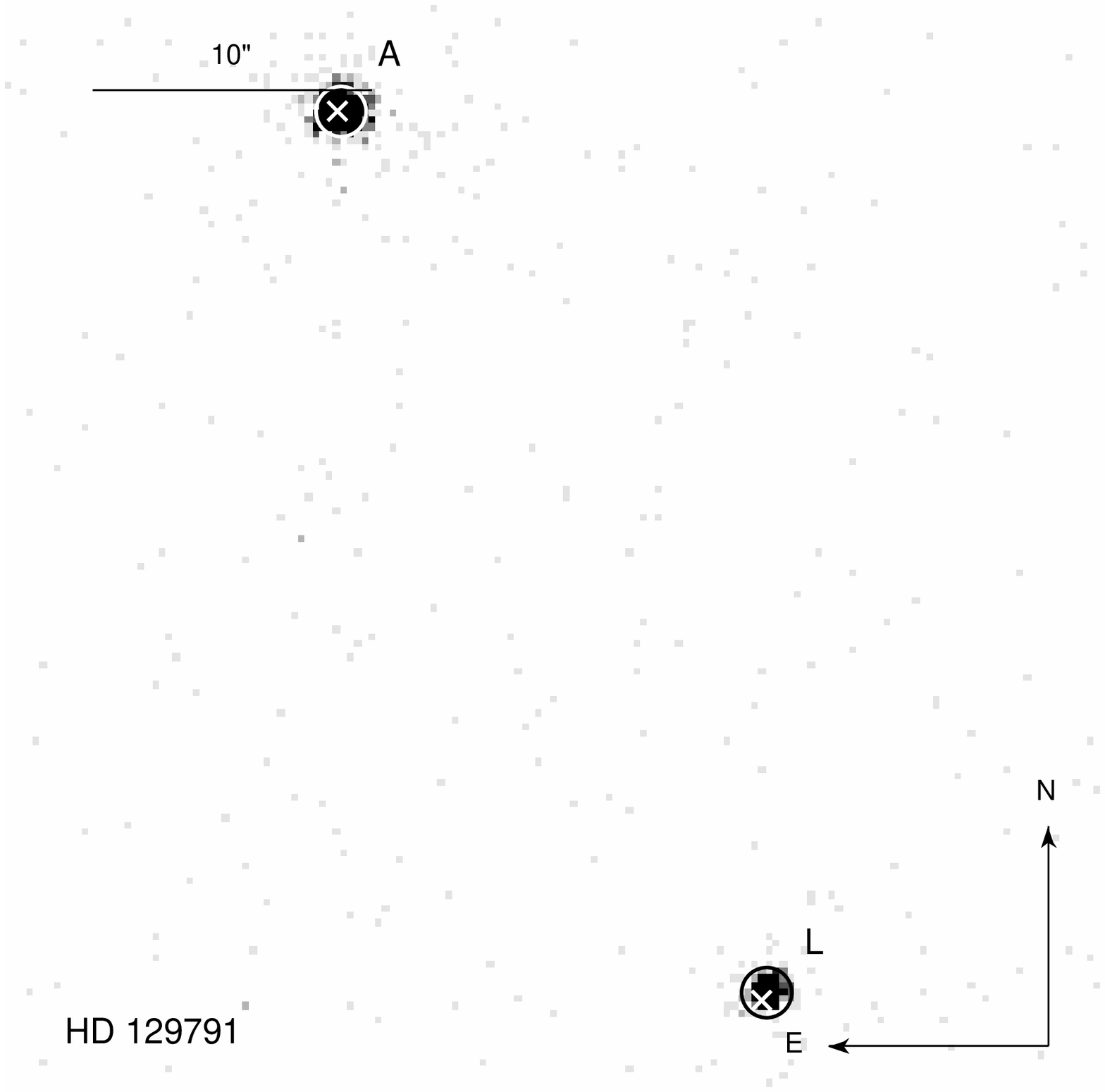}}
}
\parbox{6cm}{
\resizebox{6cm}{!}{\includegraphics{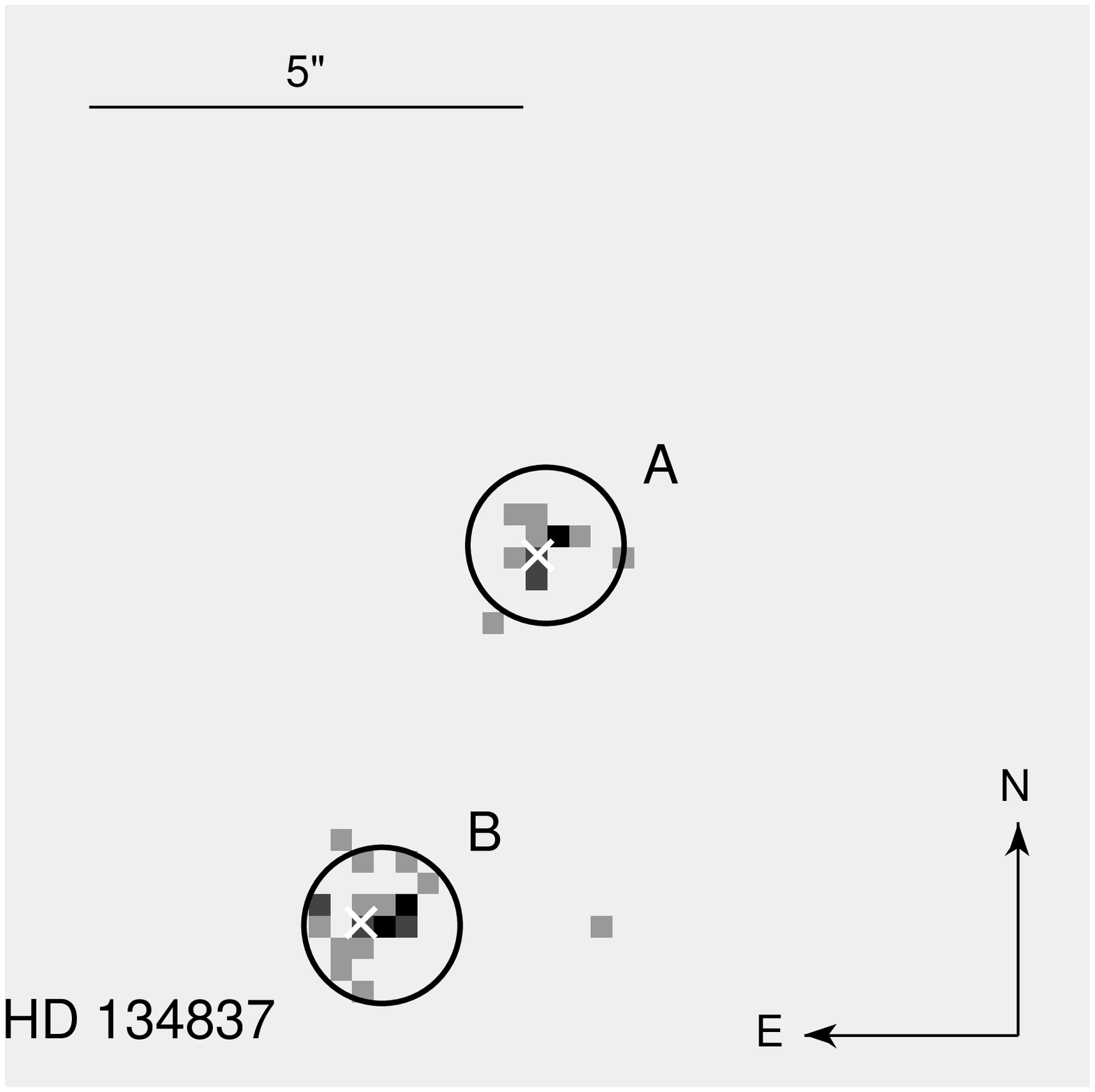}}
}
\parbox{6cm}{
\resizebox{6cm}{!}{\includegraphics{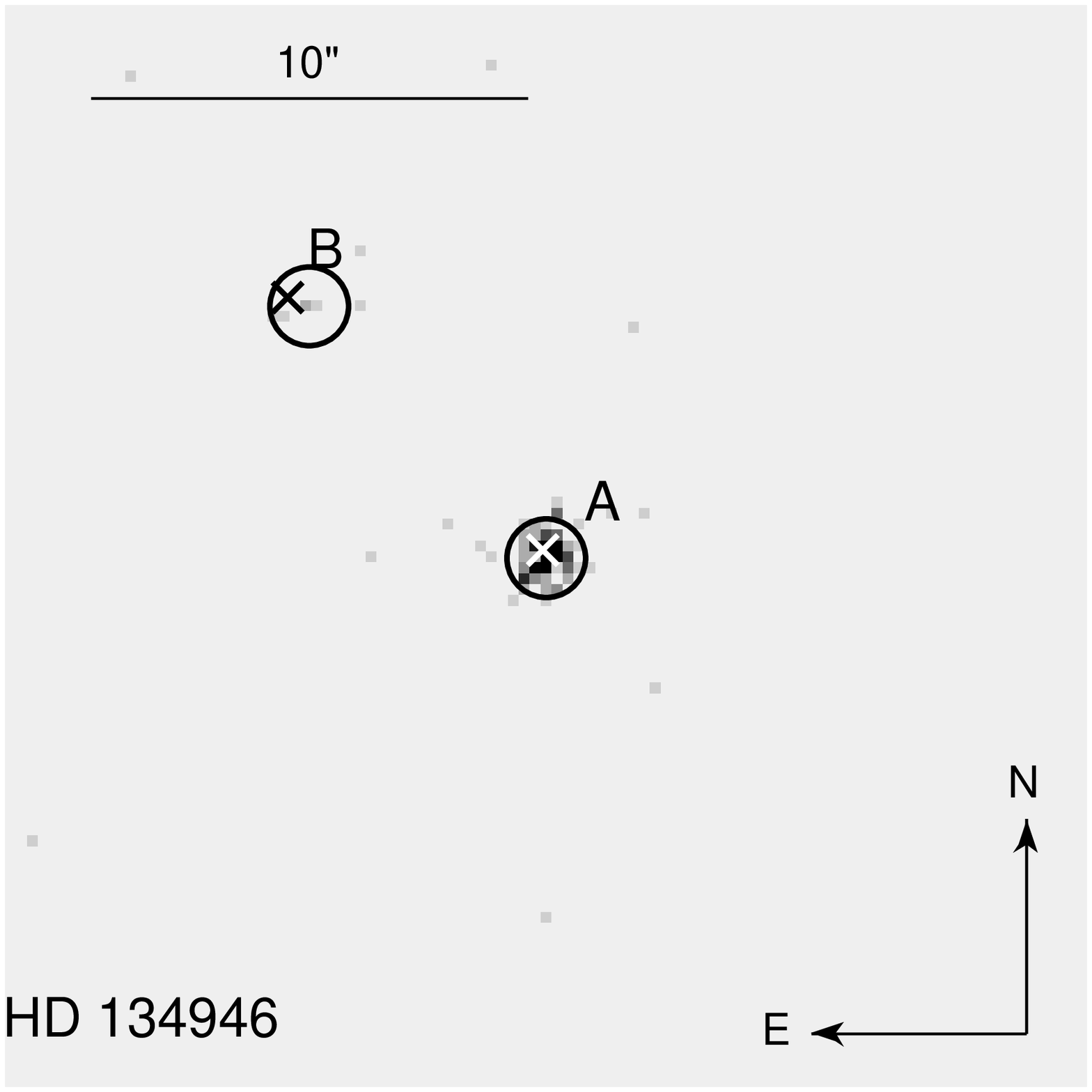}}
}
}
\caption{
{\em Chandra} ACIS images of B/A-type stars and faint companions. The size of the images is $25 \times 25$\,pix with pixel size of $0.25^{\prime\prime}$, except for HD\,129791 ($80 \times 80$\,pix) and HD\,134946 ($50 \times 50$\,pix). Crosses denote the optical/IR position of the individual components in the multiple system, circles mark the photon extraction areas centered on the position of X-ray sources detected with {\sl wavdetect}. The primaries are labeled `A', the companions with letters `B' and `L', where `L' stands for `Lindroos companion'.}
\label{fig:acis_images}
\end{center}
\end{figure*}

Fig.~\ref{fig:acis_images} shows a portion of the ACIS images 
around our targets. We overplot the photon extraction areas (circles), 
as well as the {\em Hipparcos} position of the primary and the IR position of the companions 
(x-shaped symbols).  
Surprisingly, in the new sample X-rays are detected from all but one of the primaries. 
Similar to the findings by SHH03 most companions are also X-ray emitters, 
yielding $11$ new X-ray sources, and a total of $19$ X-ray sources when
combined with the targets of SHH03. 

Table~\ref{tab:early_hrd} gives the published stellar parameters of the primaries, 
a summary of the X-ray detections, and a flag if there are
known companions that remained unresolved with {\em Chandra}. 
Among the $7$ (out of $11$) detected primaries two are known to have unresolved companions: 
HD\,169978 is a spectroscopic binary (see discussion in SHH03), and HD\,110073 is listed 
as an SB\,1 in the catalog of Ap stars by \citet{Schneider81.1}.  
However, it can not be excluded that there are close unresolved companions among the other primaries, 
because to date no systematic observations for spectroscopic companions have been carried out 
for these objects. 

Concerning the resolved components, the X-ray detection can help to single out true companions 
from chance projections. This is related to the argument laid out in the introduction,
that B-type stars on the MS are young, such that bound late-type companions are on the pre-MS where they
are naturally strong X-ray emitters. 
It is well-established that T Tauri stars have high X-ray luminosities, 
albeit with a large spread: $\log{L_{\rm x}} \sim 28...31$\,erg/s 
\citep[e.g. ][]{Stelzer01.1,Preibisch05.1}. 
For the same range of bolometric luminosity nearby field stars display much lower activity levels, 
with typically $\log{L_{\rm x}} \sim 26...28$\,erg/s \cite{Schmitt04.1}; see also Fig.~6 of
\cite{Preibisch05.1} for a direct comparison to T Tauri stars). 

Of the $15$ companions to the $11$ B-type stars observed with {\em Chandra} $12$ are detected,
and most of them show X-ray luminosities in the range $10^{29...30}$\,erg/s 
(see Table~\ref{tab:xrayparams_lx} and Sect.~\ref{subsect:xray_prop}), 
yielding strong evidence for them being pre-MS stars. However, final classification
requires confirmation of the companion status by means of spectroscopy 
(optical spectroscopy should reveal a Li\,I absorption feature at $6708$\,\AA~ 
indicating their pre-MS nature) or proper motion. 
At present this information is available for only a minority of the companions 
discussed in this article. In the meantime
available near-IR photometry allows for a rough estimate of their evolutionary stage,
comparing their position in the near-IR color-magnitude diagram (CMD) to pre-MS models.

\begin{figure}
\begin{center}
\resizebox{9cm}{!}{\includegraphics{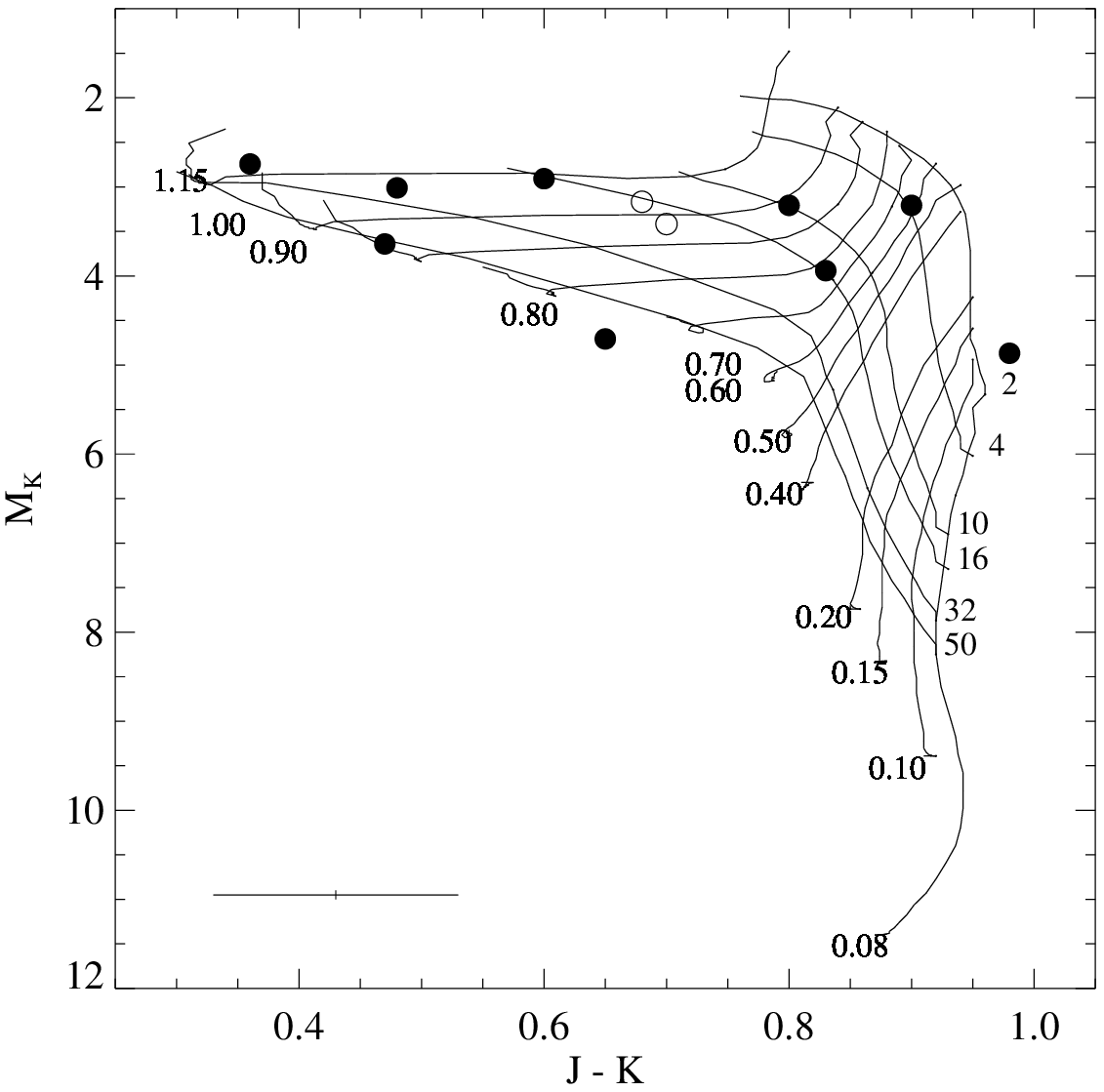}}
\caption{Near-IR color-magnitude diagram for companion candidates to intermediate-mass stars. {\em filled symbols} - objects detected with {\em Chandra}, {\em open symbols} - objects not detected with {\em Chandra}. The grid represents pre-MS models from \protect\citet{Baraffe98.1}. Errors are not shown for clarity, but are typically $0.1$\,mag (see error bar on the bottom left), such that all objects shown in this diagram are compatible with being on the pre-MS.}
\label{fig:cmd}
\end{center}
\end{figure}

Fig.~\ref{fig:cmd} shows the $M_{\rm K}$ vs. $J-K$ diagram with model calculations 
by \citet{Baraffe98.1}: $Y=0.275$, $[M/H]=0$, $\alpha_{\rm ML}=1$. 
The positions of all companion candidates with published 
$J$ and $K$ magnitudes (summarized in Table~\ref{tab:cmd}) are overplotted. 
For the distances 
needed to compute the absolute $K$-band magnitude we assumed that all companions are 
bound to the primaries. Note that three of the presumed low-mass companions discovered in
AO observations have been observed only in the $K$-band, 
such that they can not be placed in Fig.~\ref{fig:cmd}. 
For the Lindroos companions we extract the near-IR photometry from the 
2\,MASS\footnote{The 2\,MASS All Sky Catalog of point sources is available online 
at http://vizier.u-strasbg.fr/viz-bin/VizieR} catalog. 
Only one companion (HD\,1685\,B) has near-IR colors that are incompatible with the 
locus of pre-MS stars in the CMD, 
and therefore this object is likely physically unrelated to the B-type star. 
For the undetected Lindroos companions of HD\,32964 and HD\,123445, the near-IR photometry
is compatible with the pre-MS models. The upper limits for their X-ray luminosities are
between $10^{28...29}$\,erg/s, in the range where field stars and pre-MS stars overlap,
such that their nature remains unclear. 
 
We tentatively assigned a companion status to each object based on X-ray detection/non-detection 
and near-IR photometry. Corresponding flags are given in the last two columns 
of Table~\ref{tab:cmd}.
To summarize, for the majority of companions both X-rays and near-IR data consistently indicate youth. 

\begin{table}
\begin{center}
\caption{Stellar parameters for B- and A-type primaries observed with {\em Chandra}. The last two columns provide flags indicating whether the star is detected in X-rays (`X') and whether it has known companions not resolved with {\em Chandra} (`C').} 
\label{tab:early_hrd}
\begin{tabular}{lrrcrr|cc}\hline
Name       & $\log{T_{\rm eff}}$ & $\log{\frac{L_{\rm bol}}{L_\odot}}$ & Ref & vsini  & Ref & X & C \\ 
           & [K]                &                                      &     & [km/s]  &     &   &      \\
\hline
HD\,32964       & 4.05 & 2.34    & (1)   & 30  & (5) & $-$     & $\surd$ \\
HD\,73952       & 4.09 & 2.10    & (1)   &     &     & $\surd$ & \\
HD\,110073      & 4.11 & 2.59    & (1)   & 28  & (6) & $\surd$ & $\surd$ \\
HD\,129791      & 4.01 & 1.57    & (2,3) & 280 & (7) & $\surd$ & \\ 
HD\,134837      & 4.10 & 1.97    & (1)   &     &     & $\surd$ & \\
HD\,134946      & 4.14 & 2.19    & (1)   & 150 & (5) & $\surd$ & \\
\hline
HD\,1685        & 4.02 & 1.87    & (1)   & 236 & (7) & $\surd$ & \\
HD\,113703      & 4.19 & 2.67    & (4)   & 216 & (6) & $-$     & \\
HD\,123445      & 4.02 & 2.25    & (4)   &  66 & (7) & $-$     & \\
HD\,133880      & 4.08 & 2.20    & (1)   &     &     & $-$     & \\
HD\,169978      & 4.11 & 2.82    & (1)   &  91 & (6) & $\surd$ & $\surd$ \\
\hline
\multicolumn{8}{l}{(1) - \protect\citet{Hubrig01.1}, (2) - \protect\citet{deGeus89.1}, (3) - } \\
\multicolumn{8}{l}{\protect\citet{Gerbaldi01.1}, (4) - this paper: $L_{\rm bol}$ from} \\
\multicolumn{8}{l}{$V$ mag, $T_{\rm eff}$ from spectral type with conversion by} \\
\multicolumn{8}{l}{\protect\citet{Kenyon95.1}, (5) - \protect\citet{Abt02.1}, (6) - }\\
\multicolumn{8}{l}{\protect\citet{Uesugi70.1}, (7) - \protect\citet{Royer02.1}.} \\
\end{tabular}
\end{center}
\end{table}

\begin{table*}
\begin{center}
\caption{Near-IR photometry and stellar parameters for known companion candidates. 
The last two columns provide flags for the companion status:
`$\surd$' - X-ray detection or near-IR photometry suggests late-type pre-MS star and consequently likely bound system,
`?' - not resolved in X-rays or no near-IR color available.}
\label{tab:cmd}
\begin{tabular}{lrrrrrrcc}
\noalign{\smallskip}\hline\noalign{\smallskip}
HD        & \multicolumn{1}{c}{$J$}   & \multicolumn{1}{c}{$K$}   & Ref. & $\log{T_{\rm eff}}$ & $\log{\frac{L_{\rm bol}}{L_\odot}}$ & Ref & \multicolumn{2}{c}{Companionship} \\
          & \multicolumn{1}{c}{[mag]} & \multicolumn{1}{c}{[mag]} &      & [K]                 &                     &     & X-rays & NIR-phot. \\
\noalign{\smallskip}\hline\noalign{\smallskip}
\hline
HD\,32964\,L  & $8.79$  & $8.09$  & (3)  &  $3.63$ & $-0.18$ & (5)   & $-$     & $\surd$ \\
HD\,32964\,B  & $10.03$ & $9.38$  & (1)  &  $3.64$ & $-0.94$ & (1)   & $\surd$ & $\surd$ \\   
HD\,73952\,B  & $11.8$  & $10.82$ & (1)  &  $3.61$ & $-1.06$ & (1)   & $\surd$ & $\surd$ \\
HD\,129791\,L & $10.37$ & $9.54$  & (3)  &  $3.65$ & $-0.75$ & (6)   & $\surd$ & $\surd$ \\ 
HD\,110073\,B & $8.29$  & $7.93$  & (1)  &  $3.75$ & $+0.09$ & (1)   & $\surd$ & $\surd$ \\
HD\,134837\,B & $-$     & $10.99$ & (1)  &  $3.51$ & $-1.55$ & (1)   & $\surd$ & ?       \\ 
HD\,134946\,B & $-$     & $12.51$ & (1)  &  $3.43$ & $-2.18$ & (1)   & $\surd$ & ?       \\ 
\hline
HD\,113703\,B & $9.63$  & $9.16$  & (2)  &         & $-0.32$ & (2)   & $\surd$ & $\surd$ \\
HD\,113703\,L & $9.01$  & $8.53$  & (3)  &  $3.68$ & $-0.38$ & (6)   & $\surd$ & $\surd$ \\
HD\,1685\,B   & $11.7$  & $10.1$  & (1)  &  $3.59$ & $-1.20$ & (1)   & $\surd$ & $-$     \\
HD\,123445\,L & $10.54$ & $9.86$  & (3)  &  $3.65$ & $-0.81$ & (6)   & $-$     & $\surd$ \\ 
HD\,123445\,B & $10.8$  & $9.9 $  & (4)  &         & $-0.49$ & (4)   & ?       & $\surd$ \\
HD\,123445\,C & $10.7$  & $9.9$   & (4)  &         & $-0.36$ & (4)   & ?       & $\surd$ \\
HD\,133880\,B & $9.01$  & $8.41$  & (1)  &  $3.76$ & $+0.15$ & (1)   & $\surd$ & $\surd$ \\
HD\,169978\,B & $-$     & $12.69$ & (1)  &  $3.50$ & $-2.40$ & (1)   & $-$     & ?       \\
\noalign{\smallskip}\hline\noalign{\smallskip}
\multicolumn{9}{l}{(1) - \citet{Hubrig01.1}, (2) - \citet{Shatsky02.1}, (3) - \citet{Cutri03.1},} \\
\multicolumn{9}{l}{(4) - \citet{Huelamo01.1}, (5) - \citet{Lindroos85.1}, (6) - \citet{Lindroos86.1}.} \\
\end{tabular}
\end{center}
\end{table*}

\subsection{X-ray properties}\label{subsect:xray_prop}

%
\input{tab4}

In Table~\ref{tab:xrayparams_lx} we list the X-ray parameters of all known components
whether detected or not. We give the HD number of the target (column~$1$), 
component identifier (column~$2$), and a flag that distinguishes X-ray detections
(`$\surd$') from non-detections (`$-$') (column~$3$). 
The offset between X-ray and optical/IR position 
and the significance of detection resulting from the {\sl wavdetect} algorithm 
are given in cols.~$4$ and~$5$.  
The number of counts (column~$6$) refers to the $0.5-8$\,keV passband,
and comprises a fraction between $35$ and $90$\,\% of the total source counts (see col.9),
depending on the extraction radius.  
To compute upper limits for the undetected components of our target systems 
we used the method for Poisson-distributed counting data described by \citet{Kraft91.1}. 
We took account of the background 
fluctuations by estimating the background within a squared area of 
$1^\prime$ side length centered on the optical/IR position of the respective
star but excluding all detected sources, 
and scaling this mean background to the source extraction area. 
Columns~$7$ and~$8$ of Table~\ref{tab:xrayparams_lx} show hardness ratios defined as 
$HR = (H-S)/(H+S)$, where $H$ and $S$ are the number of counts in a hard band
and in a soft band, respectively.
$HR1$ is defined using emission in the $0.5-1$\,keV ($S$) and the $1-8$\,keV ($H$) bands,
and $HR2$ from the $1-2$\,keV ($S$) and the $2-8$\,keV ($H$) band.
In the final columns of Table~\ref{tab:xrayparams_lx} we list the PSF fraction included
in the source extraction area, the PSF- and absorption-corrected broad-band X-ray luminosity, 
and the ratio of X-ray to bolometric luminosity. The X-ray luminosities have been computed
with PIMMS assuming an iso-thermal emitting plasma with $kT=1$\,keV. Absorption is
neglected, except for HD\,129791 where a column density of 
$N_{\rm H} = 5.0 \times 10^{20}\,{\rm cm^{-2}}$ 
is assumed, corresponding to the value derived from the optical 
extinction \citep[$A_{\rm V} = 0.26$\,mag; ][]{Ryter96.1}.  
In general stellar coronae are characterized by a multi-temperature plasma. Therefore, 
the assumption of a 1-T model may not be appropriate, and consequently
the X-ray luminosities may not be very precise. 
In Sect.~\ref{subsect:spectra} more reliable X-ray luminosities are derived directly
from the X-ray spectrum for the brighter sources.

\subsection{X-ray lightcurves}\label{subsect:lcs}

\begin{figure}
\begin{center}
\resizebox{9cm}{!}{\includegraphics{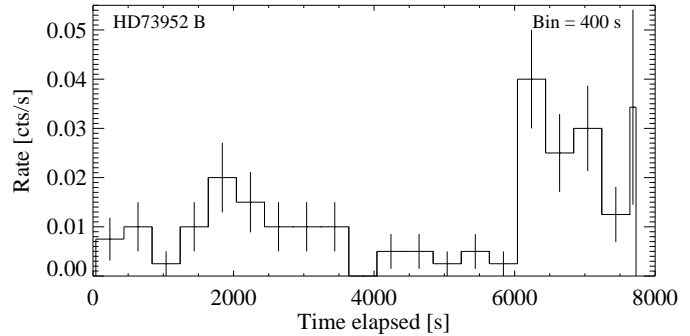}}
\caption{{\em Chandra} ACIS lightcurve of HD\,73952\,B, the only significantly variable star in the sample.}
\label{fig:acis_lcs}
\end{center}
\end{figure}

We binned lightcurves for all detected sources in $400$\,s intervals.
Obviously, the detection of variability requires both high statistics and a good time
resolution, and our choice of the binning is a compromise between these two
opposing effects. 
The duration of these snapshop observations ($\sim 2-12$\,ksec) 
is well below typical variability time-scales on active stars, and therefore we do not 
expect to detect a lot of variability. Indeed, a KS-test yields only one of the
$19$ X-ray sources in the sample variable at $>99$\,\% confidence level (HD\,73952\,B). 
The lightcurve of this object is displayed in Fig.~\ref{fig:acis_lcs}, with a sharp
increase in the count rate about two thirds into the observation. 

The occurrence of $1$ flare in $\sim 80$\,ksec (the sum of the observing time of all
detected stars in the total sample) is not unusual.
Actually, it is even above the expectation. 
\citet{Wolk05.1} derived a rate of $1$ flare per $640$\,ksec from a detailed
variability analysis of young solar-analogs in Orion. 
However, due to the observational limitations in our short snapshots we do not give a 
lot of weight to this difference, and, more generally, with this data set variability 
can not be used to examine the origin of the X-rays. 

\subsection{X-ray spectra}\label{subsect:spectra}

%
%
\input{tab5}

%
\begin{figure}
\begin{center}
\resizebox{9cm}{!}{\includegraphics{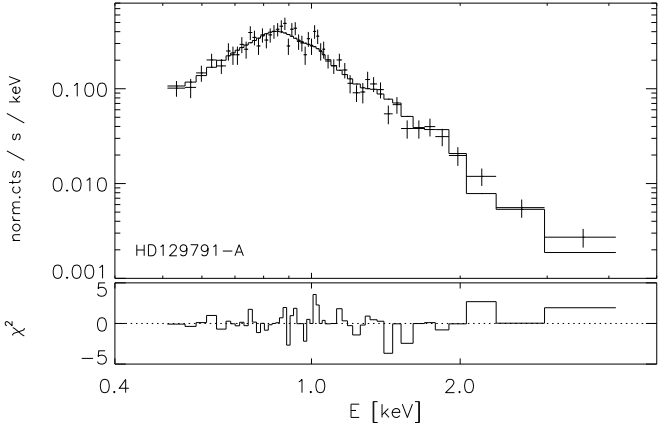}}
\caption{X-ray spectrum of HD\,129791\,A: data, best fit model, and residuals. The 2-T fit requires an absorption term with 
$N_{\rm H} = 1.7 \times 10^{21}\,{\rm cm^{-2}}$.}
\label{fig:acis_spec}
\end{center}
\end{figure}

For most of the X-ray sources in the sample presented here the number of counts collected 
is small but sufficient for a basic description of the temperature and luminosity 
of the emitting plasma.
As shown by \citet{Feigelson02.1} for data of similar quality the typical uncertainties in $kT$ 
for sources between $30$ and $100$ counts with ACIS are about $30...60$\,\%. 
Thus, although limited in detailed information, the derived parameters are not
meaningless. In our earlier study (SHH03) we have shown the improvement
in the estimate of the spectral parameters obtained with similar quality ACIS spectra
versus earlier observations of {\em ROSAT}.

For all but the faintest X-ray sources from Table~\ref{tab:xrayparams_lx} 
we extracted a spectrum, the corresponding detector response matrix that maps
pulse heights into energy space, and an
auxiliary response file which contains information about the effective area
and detector efficiency across the chip as a function of energy.
We binned each spectrum to a minimum of $5$ or more counts per bin starting at
$0.5$\,keV.  
As the background of ACIS is very low ($< 1$\,count in the source extraction area)
it can be neglected. 
Spectral modelling was performed in the XSPEC environment, version 11.3.0. 

First we approximated each spectrum 
with a one-temperature (1-T) thermal model 
(APEC\footnote{For a description of the Astrophysical Plasma Emission Code ({\sc APEC}) see \cite{Smith01.1}.}).
Some of the spectra have a high-energy excess with respect to 
the 1-T model. These spectra are better described when a second component with higher
temperature is added, as verified by a significant reduction of $\chi^2_{\rm red}$ as 
well as visual inspection. Since the hot component is required only for the brightest sources, 
its apparent absence in the remainder of the spectra can likely be attributed to the
poorer signal at high energies, rather than being a true physical difference. 
A direct comparison shows that for most spectra the temperature from the 1-T model agrees 
reasonably well with the mean temperature of the 2-T model, defined as the average
of the two temperatures weighted with the emission measures. 
This test yields a justification for the wide-spread use of 1-T models in cases of poor statistics.
However, to obtain a consistent description of all spectra, we choose always the 2-T model as best fit 
(see Table~\ref{tab:xrayspectab}). 

Extinction is likely to be negligible, because the stars are evolved enough such
that no substantial amount of circumstellar material is expected to be present. 
In fact, HD\,129791 is the only star with measured non-vanishing optical extinction. 
To test the role of absorption in the X-ray spectra we include a photo-absorption term 
comprising the atomic cross-section and elemental abundances from \citet{Wilms00.1}. 
Two sets of spectral fits were performed, one with free column density ($N_{\rm H}$) and one
with $N_{\rm H}$ set to zero. 
The results show that mostly the model without absorption is adequate, i.e. the model
with free column density either yields $N_{\rm H} = 0$ or the fit is poorer 
(based on the $\chi^2$ statistics and visual inspection of the residuals). 
An exception is HD\,129791\,A. Here, a photo-absorption term is
required to describe the spectral shape below $1$\,keV (see Fig.~\ref{fig:acis_spec}). 
HD\,133880\,B and HD\,113703\,B may be better described with a 3-T model -- see SHH03 for a 
graphical representation of their X-ray spectra -- but the highest temperature
does not make a large contribution to the total emission measure, such that we can
safely ignore it.  

The luminosities in Table~\ref{tab:xrayspectab} refer to the $0.5-8.0$\,keV broad-band, 
after correction for absorption in the case of HD\,129791\,A. 
These luminosities have a tendency of being slightly larger than the values derived with PIMMS 
assuming an iso-thermal plasma of temperature $1$\,keV.
The emission measure weighted mean temperatures extracted from the spectra are generally 
slightly smaller than 
$1$\,keV. But this is expected to produce an opposite trend, i.e. $L_{\rm x}$ is rather expected to be
smaller. However, the discrepancy is only $0.1-0.2$\,dex in $\log{L_{\rm x}}$, and may
be ascribed to the use of different plasma codes: RS \citep{Raymond77.1} 
within PIMMS and APEC for the spectral modeling.

\section{The $L_{\rm x}/L_{\rm bol} - $relation}\label{sect:lxlbol}

\begin{figure}
\begin{center}
\resizebox{9cm}{!}{\includegraphics{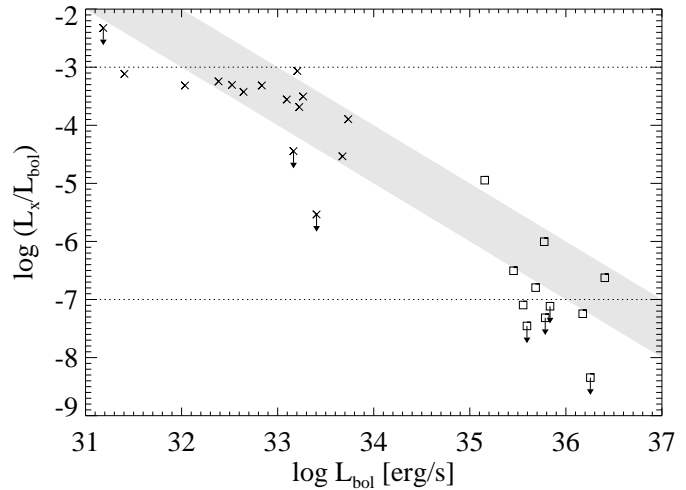}}
\caption{Ratio between X-ray and bolometric luminosity for A- and B-type stars (squares) and their companions (x-points). 
The shaded area indicates the position of objects in the range $L_{\rm x} = 10^{29...30}$\,erg/s. 
Dotted lines indicate the canonical value for $L_{\rm x}/L_{\rm bol}$ of late-type stars ($10^{-3}$) and of hot stars ($10^{-7}$). 
For the unresolved pair of companions HD\,123445\,B and~C it was assumed that the observed X-ray luminosity is distributed equally on both stars.}
\label{fig:acis_lx_lbol}
\end{center}
\end{figure}

Fig.~\ref{fig:acis_lx_lbol} displays the $L_{\rm x}/L_{\rm bol}$ ratio 
for all components. In the sample
investigated here the companions display $\log{(L_{\rm x}/L_{\rm bol})}$ 
values near the saturation limit. The spread is remarkably small,
compared to the $2-3$\,dex observed typically for late-type pre-MS stars.  
Judging from the age of the primaries (on the order of $10-100$\,Myr),
the companions are probably relatively evolved and non-accreting.
The absence of accretors diminuishes the scatter of $L_{\rm x}/L_{\rm bol}$
in a sample of pre-MS stars because the latter ones have systematically lower
X-ray luminosities; see e.g. \citet{Preibisch05.1}. 

For all intermediate-mass stars of the sample which are not detected the upper
limits we derive are lower than the canonical value of $10^{-7}$ for hot stars.
Although only one of the four non-detections is clearly below this value, 
we conjecture that none of these sources are wind-driven.  
The detected B/A stars show values of $\log{(L_{\rm x}/L_{\rm bol})}$,
which are somewhat higher than expected for `classical' wind X-ray sources.  
However, we caution that recent
studies have demonstrated that not all hot stars fullfill the 
$\log{(L_{\rm x}/L_{\rm bol})} = 10^{-7}$ relation \citep{Stelzer05.2}. 

In terms of absolute X-ray luminosity the detected intermediate-mass stars and the companions are
indistinguishable, with the majority showing $L_{\rm x} \sim 10^{29...30}$\,erg/s.
This is indicated by the shaded area in Fig.~\ref{fig:acis_lx_lbol}.  
Assuming that the X-ray emission from the B/A stars is produced by as yet 
unknown close cool magnetically active companions which emit at typical levels 
of $\log{(L_{\rm x}/L_{\rm bol})} \sim -3.5$,
we find bolometric luminosities between $\log{(L_{\rm bol}/L_\odot}) \sim 0.0 ... -1.5$ 
for such objects. As can be seen from Table~\ref{tab:cmd} 
this is exactly the range of $L_{\rm bol}$ for 
the companions that have already been identified and spatially resolved. 

\section{Discussion}\label{sect:discussion}

In this paper we have presented the {\em Chandra} observations of $6$ late-B 
stars on the MS. The sample complements a previous survey of $5$ B-type stars presented by SHH03. 
The targets are characterized by (1) being known to emit X-rays from {\em ROSAT}, 
and (2) having close companions that could not be spatially separated in previous X-ray images, 
leaving doubt on the origin of their {\em ROSAT} detection. 
We find that the B-type primary is detected in $7$ out of $11$ cases, after resolving it
from all known visual companions with {\em Chandra}.  
A trend of the early-type 
stars to split in two groups was pointed out by \citet{Daniel02.1} for the 
Pleiades: Apparent X-ray emitters on the one side, 
and on the other side X-ray quiet stars with upper limits by $1-2$ orders 
of magnitudes lower than the $L_{\rm x}/L_{\rm bol}$ values of the detected B-type stars. 
Our observations are consistent with this general picture, although the two populations
are not clearly separated. 
The two possible interpretations for the observed behavior are 
(i) X-ray emission is intrinsic to some B-type stars, 
or (ii) the detected late-B type stars have further (as yet undiscovered) low-mass companions. 

The detection fraction of $> 60$\,\% for the primary stars is surprisingly high. 
We confront these numbers with the sample of RASS detected late-B type stars.
The catalog by BSC96 counts $121$ X-ray detected stars with spectral types between B5...B9
versus $973$ non-detections in the same spectral type range (12\,\% detection rate). 
A systematic search of the literature revealed that for $76$ 
among the $121$ RASS detections closeby fainter companions are known, that are confused 
with the B-type star in the {\em ROSAT} error box.  
The {\em Chandra} targets are drawn from this sample. Therefore, it seems justified to 
assume that our B star detection rate of $> 60$\,\% is applicable to all known visual
late-B type binaries. If the residual X-ray emission at the primary position is attributed
to unknown companions, this implies then that about $(76/121) \times (7/11) \sim 40$\,\% 
of X-ray detected B-type stars are high 
multiplicity systems. This seems at first hand unlikely. However, a speckle survey 
has revealed a fraction of $\sim 50$\,\% of high-order multiples among known spectroscopic binaries 
of HgMn type \citep{Isobe91.1}. Furthermore, the near-IR studies by \citet{Hubrig05.1} have brought forth the
discovery of third components in many SB systems among chemically peculiar (CP) stars.

In the {\em Chandra} sample three stars are known to have CP character: HD\,32964 and HD\,110073 
are HgMn stars, 
and both are spectroscopic binaries, i.e. taking into account their visual companions 
both systems are probably at least triples. 
Interestingly, HD\,32964 is composed of two nearly
equal mass B-type stars, that apparently both are X-ray dark, because the only X-ray source is
associated with the visual companion. HD\,110073, is detected with {\em Chandra}, but 
it might have a magnetic field (see below), suggestive of X-ray production on the intermediate-mass 
star itself. 
The third CP star is HD\,133880, an Ap star which has no known spectroscopic companion,
and it is also not detected in X-rays. 

As an alternative to the companion hypothesis, 
the high X-ray detection fraction of the primaries may indicate that --
contrary to the common believe -- at least some of them are capable to produce high-energy emission. 
Intermediate-mass stars possess neither strong winds nor a solar-type dynamo, but they may 
represent a hybrid case. A magnetically confined wind has been proposed 
by \citet{Babel97.1} to explain the unusually strong X-ray luminosity 
from the magnetic Ap star IQ\,Aur. 
In contrast to the classical wind-shock picture, this model can also produce relatively 
hard X-ray emission, with temperatures in excess of $10^7$\,K produced in shocks that form upon 
collision of the magnetically channeled winds from the two stellar hemispheres. The high observed X-ray 
temperatures of the {\em Chandra} sources identified with the primaries ($\sim 10$\,MK) therefore do 
not exclude that the emission comes from the B stars. 
But, obviously, the general applicability of this scenario 
to A- and B-type stars depends on the presence of magnetic fields. 
Their detection is difficult in intermediate-mass
stars because they have fewer absorption lines and faster rotation rates than lower-mass stars. 
\citet{Bychkov03.1} have compiled a list of field measurements for $\sim 600$ MS 
and giant stars. This catalog comprises $22$ of the $121$ X-ray emitting late-B stars
from BSC96, of which $8$ are apparently single stars. 

Two stars of the {\em Chandra} sample 
are listed by \citet{Bychkov03.1}, but they behave contrary to the expectation: 
The field measurement for HD\,110073 is hardly significant. 
This star is detected with {\em Chandra}, but recall that it is a spectroscopic binary. 
HD\,133880 is an Ap star with a strong, probably non-dipolar field 
\citep[$2-4$\,kG; ][]{Landstreet90.1, Bychkov03.1},  
but in SHH03 we have shown that the X-ray emission can be attributed to the companion. 
Therefore, the presence of magnetic fields does not necessarily imply a relation to X-ray emission. 

The short exposure times of the {\em Chandra} observations 
do not allow to examine X-ray properties of the detected sources in detail.
Therefore, it is difficult to tell if the X-ray emission from the position of the primaries 
differs qualitatively from the X-rays produced by the resolved companions. 
Comparison of the X-ray luminosities and temperatures for the primaries on the one hand  
and the companions on the other hand does not indicate any systematic trend 
(cf. Table~\ref{tab:xrayspectab} and Fig.~\ref{fig:acis_lx_lbol}). 
This is consistent with the same emission mechanism working in both groups. 
Collectively, the X-ray luminosities and temperatures of the objects studied 
in this paper are much lower than those of low-mass pre-MS stars in star forming 
regions \citep[e.g. ][]{Preibisch02.1, Getman02.1, Feigelson02.1}. 
When choosing energies of $0.5-2.0$\,keV for the soft band
and $2.0-8.0$\,keV for the hard band, as common for pre-MS objects,
all our targets show very negative hardness ratios ($HR = -0.8$ and below),
indicating that no significant emission is present at high energies.  
Thus, both spectral hardness and X-ray luminosity put our targets inbetween the Pleiades and the bulk
of pre-MS stars. If this is to be interpreted as an evolutionary effect the low-mass
stars in the sample presented here should be in their final approach to the MS,
in agreement with their evolutionary state derived from the IR magnitudes and
colors.  

The fact that nearly all known resolved companion candidates are shown to be X-ray sources,  
and the similarity of their X-ray properties with those of the B-type primaries, 
motivates us to speculate about the potential of such objects in eventually explaining the
observed X-ray emission from all or most MS B-type stars. 
In this vein, X-ray detections could even be used as a tool to discover faint companions to B-type stars.
The most likely exceptions are the Ap/Bp stars. Checking whether they form a class of intrinsic 
X-ray emitters must include a systematic study of their multiplicity. 
In our future work we will continue to 
examine the multiplicity of the apparently X-ray active B-type stars through IR
imaging \citep[see ][for preliminary results]{Stelzer05.1, Huelamo05.1}, 
and IR/optical spectroscopy. Further high resolution X-ray observations are also desirable. 
Combining these techniques on a large sample is essential to solve the longstanding mystery
of the X-rays from intermediate-mass stars.

\appendix

\section{Individual Targets}\label{sect:newdata}

\subsection{HD\,32964}\label{subsect:hd32964}

HD\,32964 ( = 66\,Eri) is a CP star of HgMn type \citep{Renson91.1}, 
and a spectroscopic binary \citep{Frost24.1}
composed of two nearly equal mass stars of $M \sin^3{i} \sim 2.4\,{\rm M_\odot}$ 
in a $5.2$-day orbit \citep{Yushchenko01.1, Catanzaro04.1}.

HD\,32964 is also a Lindroos system with a secondary at separation of 
$\sim 53^{\prime\prime}$. \citet{Lindroos85.1} noted that common proper motion 
is observed, but \citet{Eggen63.1} classified the K5\,V secondary as optical, 
and based on this classification it was further on discarded from the Lindroos sample. 
Consequently, this star was not included in the X-ray study of \citet{Huelamo00.1}.
In the RASS an X-ray source was detected with $\log{L_{\rm x}} = 29.8$\,erg/s 
\citep{Berghoefer96.1}.
HD\,32964 was also part of the {\em ROSAT} HRI sample examined by \citet{Berghoefer94.1}.
While the Lindroos secondary remained undetected consistent with its classification
as non-physical companion, the primary was detected at a level of 
$L_{\rm x} = 7 \times 10^{29}$\,erg/s. 

After the {\em ROSAT} study another, presumably late-type, companion to HD\,32964 
was discovered by \citet{Hubrig01.1} with AO observations. 
Near-IR photometry puts the object near the zero-age MS in the CMD 
(Fig.~\ref{fig:cmd}), in agreement with the relatively old age ($\sim 200$\,Myrs)
derived for the primary by \citet{Hubrig01.1}. 
With a separation of 
only $\sim 1.6^{\prime\prime}$ this object was not resolvable from the primary with the
{\em ROSAT} HRI. 
The only X-ray source detected with {\em Chandra} in the vicinity of HD\,32964 is clearly
identified with this companion. 
The upper limit for the B-type primary is $L_{\rm x} < 1.6 \times 10^{28}\,{\rm erg/s}$.
The Lindroos secondary remains again undetected at an upper limit of $L_{\rm x} < 8 \times 10^{27}\,{\rm erg/s}$,
yielding further evidence that it is not a young late-type star, 
despite it being compatible with the pre-MS tracks.

\subsection{HD\,73952}\label{subsect:hd73952}

HD\,73952 is a member of the $30-50$\,Myr-old open cluster IC\,2391 \citep{Hogg60.1, Dodd04.1}.
\citet{Patten93.1} reported the X-ray detection of HD\,73952 in a {\em ROSAT}
pointed observation, with $L_{\rm x}$ four times overluminous according to the
canonical relation of $L_{\rm x}/L_{\rm bol}=10^{-7}$ for early-type stars.
\citet{Berghoefer96.1} list HD\,73952 with an X-ray luminosity of 
$\log{L_{\rm x}}=29.7\,{\rm erg/s}$. 

HD\,73952 is not a RV variable according to \citet{Andersen83.1}.
But recent AO observations revealed a faint IR object at $1.2^{\prime\prime}$ from 
the B-type star \citep{Hubrig01.1}. Its position in the $JHK$ CMD suggests that it
is a very young star of very low mass.

The {\em Chandra} image shows two X-ray sources coincident with either of the two
components in the HD\,73952 system. The brighter source is associated with the secondary,
and displays a typical `late-type' $L_{\rm x}/L_{\rm bol}$ ratio. 
For the B-star we measure $\log{(L_{\rm x}/L_{\rm bol})} = -6.7$, clearly -- and not surprisingly -- 
lower than indicated by {\em ROSAT}.

\subsection{HD\,110073}\label{subsect:hd110073}

HD\,110073 is a CP star of the HgMn class \citep{Renson91.1}.
An averaged quadratic magnetic field of $145 \pm 158$\,G was reported \citep{Bychkov03.1}
combining earlier measurements from the literature. As can be seen from the large
$1\sigma$ uncertainty the detection of the magnetic field is highly uncertain.
A faint IR object at a separation of $1.2^{\prime\prime}$ 
seems to be a comparatively massive $> 1\,M_\odot$ star on the pre-MS \citep{Hubrig01.1}.
Furthermore, HD\,110073 is a single-lined spectroscopic binary \citep{Schneider81.1}. 

Two close X-ray sources are detected with {\em Chandra}, 
spatially coincident with the optical/IR position of HD\,110073\,A and the companion
discovered in the AO observations. The presence of the spectroscopic companion
casts some doubt on the origin of the X-rays from the position of the primary,
while the possible detection of a magnetic field makes it an interesting candidate
for being an intrinsic B-star X-ray emitter.

\subsection{HD\,134837}\label{subsect:hd134837}

HD\,134837 is a member of the Upper Centaurus Lupus (UCL) association
\citep{deZeeuw99.1}. It was not known to be a multiple prior to the AO survey 
by \citet{Hubrig01.1}, which revealed an IR object at $4.7^{\prime\prime}$ from 
the B-star. The same object was recovered in a recent AO study of the Sco\,OB
association by \citet{Kouwenhoven05.1}.

In the {\em Chandra} image both the primary and the IR object are detected
as faint sources.

\subsection{HD\,134946}\label{subsect:hd134946}

HD\,134946 is a poorly studied late-B type star for which \citet{Hubrig01.1}
identified a possible companion at $8.2^{\prime\prime}$ separation. 
In our {\em Chandra} snapshot both objects are detected, albeit the companion
only marginally ($5$ photons corresponding to $\log{L_{\rm x}} = 28.3$\,erg/s). 
The primary is the brighter X-ray source with an $L_{\rm x}/L_{\rm bol}$ ratio
of $\sim -6$.

\subsection{HD\,129791}\label{subsect:hd129791}

HD\,129791 is a member of the UCL association, 
and a Lindroos system \citep{Lindroos83.1}. 
According to \citet{Gahm83.1} the secondary has spectral type K5\,V,
and in a spectroscopic study by \citet{Pallavicini92.1} it showed all
signs of youth (Ca\,II and H$\alpha$ emission, strong Lithium feature).
Its separation from the B9\,V primary is $\sim 35^{\prime\prime}$. 
Both the primary and the secondary of this system were detected with {\em ROSAT}
\citep{Huelamo00.1}. This led \citet{Huelamo01.1} to search for a further, closer 
companion to HD\,129791\,A with AO imaging, but the result was negative. 
Another companion search using AO by \citet{Kouwenhoven05.1} resulted also 
negative. 

The $\sim 7$\,ksec long {\em Chandra} exposure confirms the detection of both 
components, but no other X-ray source near the primary, consistent with the
absence of further companions in the IR. The primary of HD\,129791 is the
X-ray brightest of the B-star sample, both in absolute numbers and in terms
of the fractional X-ray luminosity ($\log{(L_{\rm x}/L_{\rm bol})} \sim -5$).

\begin{acknowledgements}
Very special thanks to Kevin Briggs for original thoughts 
that we tried to incorporate in this work. 
We would like to thank the anonymous referee for useful comments. 
This publication makes use of data products from the Two Micron All Sky Survey,
 which is a joint project of the University of Massachusetts and the Infrared 
Processing and Analysis Center/California Institute of Technology, 
funded by the National Aeronautics and Space Administration and the National Science Foundation.
This research has made use of the SIMBAD database, operated at CDS, Strasbourg, France,
and the {\em Hipparcos} catalogue accessed through the VizieR data base. 
\end{acknowledgements}

\end{document}

%% file: tab4.tex
\begin{table*}\begin{center}
\caption{X-ray parameters of all components in the sample; see text in Sect.3.3.}
\label{tab:xrayparams_lx}
\begin{tabular}{lccrrrrrrrrrr}\hline
Designation & Opt/IR & X-rays & $\Delta_{\rm xo}$                       & Sign. & Counts$^*$ & \multicolumn{1}{c}{HR\,1} & \multicolumn{1}{c}{HR\,2} & PSF & $\log{L_{\rm x}^*}$ & $\log{(\frac{L_{\rm x}^*}{L_{\rm bol}})}$ \\
            &        &        & \multicolumn{1}{c}{[$^{\prime\prime}$]} &       &            &                           &                           & [\%]      & [erg/s]             &                                   \\ \hline
HD\,32964 & A & $-$     & $-$ & $-$ & $ <     8.2$ & $-$ & $-$ &  0.88 & $<  28.2$ & $<  -7.4$ \\
          & B & $\surd$ & $   0.20 $ & $  43.9 $ & $   99.0 \pm   11.0$ & $  0.15 \pm   0.16$ & $ -0.72 \pm   0.22$ & $ 0.88$ & $  29.3$ & $  -3.4$ \\
          & L & $-$     & $-$ & $-$ & $ <     4.4$ & $-$ & $-$ &  0.90 & $<  27.9$ & $<  -5.5$ \\
HD\,73952 & A & $\surd$ & $   0.23 $ & $  16.1 $ & $   43.0 \pm    7.6$ & $ -0.02 \pm   0.27$ & $ -0.90 \pm   0.38$  & $ 0.83$ & $  29.0$ & $  -6.7$ \\
          & B & $\surd$ & $   0.19 $ & $  39.0 $ & $   92.0 \pm   10.6$ & $  0.22 \pm   0.17$ & $ -0.54 \pm   0.23$ & $ 0.83$ & $  29.3$ & $  -3.2$ \\
HD\,110073 & A & $\surd$ & $   0.13 $ & $  16.0 $ & $   28.0 \pm    6.4$ & $ -0.21 \pm   0.35$ & $ -0.45 \pm   0.68$ & $ 0.77$ & $  29.0$ & $  -7.1$ \\
           & B & $\surd$ & $   0.08 $ & $  20.5 $ & $   45.0 \pm    7.8$ & $ -0.02 \pm   0.26$ & $ -0.91 \pm   0.37$ & $ 0.77$ & $  29.3$ & $  -4.4$ \\
HD\,134837 & A & $\surd$ & $   0.15 $ & $   6.2 $ & $   12.0 \pm    4.6$ & $  0.00 \pm   0.60$ & $ -0.67 \pm   1.02$ & $ 0.90$ & $  28.5$ & $  -7.0$ \\
           & B & $\surd$ & $   0.25 $ & $   8.1 $ & $   22.0 \pm    5.8$ & $  0.00 \pm   0.40$ & $ -0.64 \pm   0.65$ & $ 0.90$ & $  28.8$ & $  -3.3$ \\
HD\,134946 & A & $\surd$ & $   0.20 $ & $  57.2 $ & $  152.0 \pm   13.4$ & $  0.13 \pm   0.13$ & $ -0.81 \pm   0.17$ & $ 0.90$ & $  29.8$ & $  -6.0$ \\
           & B & $\surd$ & $   0.53 $ & $   1.6 $ & $    5.0 \pm    3.4$ & $ -0.20 \pm   1.17$ & $ -1.00 \pm   3.09$ & $ 0.90$ & $  28.3$ & $  -3.1$ \\
HD\,129791 & A & $\surd$ & $   0.13 $ & $ 236.6 $ & $ 1471.0 \pm   39.4$ & $ -0.14 \pm   0.04$ & $ -0.77 \pm   0.06$ & $ 0.90$ & $  30.3$ & $  -4.9$ \\
           & L & $\surd$ & $   0.33 $ & $ 101.2 $ & $  305.0 \pm   18.5$ & $ -0.30 \pm   0.08$ & $ -0.85 \pm   0.14$ & $ 0.90$ & $  29.6$ & $  -3.3$ \\
\hline
HD\,113703 & A & $-$     & $-$ & $-$ & $ <    17.5$ & $-$ & $-$ &  0.68 & $<  28.1$ & $<  -8.2$ \\
           & B & $\surd$ & $   0.21 $ & $ 228.4 $ & $ 1257.0 \pm   36.5$ & $ -0.34 \pm   0.04$ & $ -0.77 \pm   0.07$ & $ 0.90$ & $  29.8$ & $  -3.5$ \\
           & L & $\surd$ & $   0.43 $ & $ 341.2 $ & $ 2943.0 \pm   55.3$ & $ -0.12 \pm   0.03$ & $ -0.67 \pm   0.04$ & $ 0.90$ & $  30.2$ & $  -3.0$ \\
HD\,1685 & A & $\surd$ & $   0.20 $ & $  17.1 $ & $   42.0 \pm    7.5$ & $  0.00 \pm   0.27$ & $ -0.90 \pm   0.38$ & $ 0.90$ & $  29.0$ & $  -6.5$ \\
         & B & $\surd$ & $   0.29 $ & $  27.7 $ & $   66.0 \pm    9.2$ & $  0.03 \pm   0.21$ & $ -0.65 \pm   0.31$ & $ 0.90$ & $  29.2$ & $  -3.2$ \\
         & $-$ & $\surd$ & $    - $ & $   2.6 $ & $    5.0 \pm    3.4$ & $  0.20 \pm   1.17$ & $  0.33 \pm   2.00$ & $ 0.90$ & $  28.1$ & $   -  $ \\
HD\,123445 & A & $-$     & $-$ & $-$ & $ <     4.4$ & $-$ & $-$ &  0.90 & $<  28.8$ & $<  -7.1$ \\
           & B & $\surd$ & $   0.49 $ & $  27.2 $ & $   59.0 \pm    8.7$ & $  0.05 \pm   0.22$ & $ -0.81 \pm   0.31$ & $ 0.90$ & $  29.9$ & $  -3.2$ \\
           & C & ${\prime\prime}$  & $   0.46 $ & ${\prime\prime}$ & ${\prime\prime}$ & ${\prime\prime}$ & ${\prime\prime}$ & ${\prime\prime}$ & ${\prime\prime}$ & $  -3.3$ \\
           & L & $-$ & $-$ & $-$ & $ <     4.4$ & $-$ & $-$ &  0.90 & $<  28.8$ & $<  -4.4$ \\
HD\,133880 & A & $-$     & $-$ & $-$ & $ <     8.2$ & $-$ & $-$ &  0.35 & $<  28.9$ & $<  -6.9$ \\
           & B & $\surd$ & $   0.24 $ & $  64.2 $ & $  193.0 \pm   14.9$ & $  0.16 \pm   0.11$ & $ -0.55 \pm   0.15$ & $ 0.90$ & $  29.9$ & $  -3.8$ \\
HD\,169978 & A & $\surd$ & $   0.15 $ & $  47.1 $ & $  122.0 \pm   12.1$ & $  0.02 \pm   0.15$ & $ -0.65 \pm   0.21$ & $ 0.90$ & $  29.8$ & $  -6.6$ \\
           & B & $-$     & $-$ & $-$ & $ <    14.6$ & $-$ & $-$ &  0.90 & $<  28.9$ & $<  -2.3$ \\
\hline
\multicolumn{11}{l}{$^*$ in the $0.5-8$\,keV passband; $L_{\rm x}$ refers to the distance given in Table~\ref{tab:obslog} and has been corrected from the encircled PSF fraction.} \\
\end{tabular}
\end{center}\end{table*}

%% file: tab5.tex
\begin{table*}\begin{center}
\caption{Spectral parameters of X-ray sources associated with {\bf B/A stars} 
and/or their known close companions.
The absorption is negligible, except for HD\,129791\,A. The higher temperature of HD\,1685\,B is 
unconstrained and the $90$\,\% confidence lower limit is given.}
\label{tab:xrayspectab}
\begin{tabular}{lcrrrrrrr}\hline
HD & Opt/IR & $\chi^2_{\rm red}$ (dof) & $\log{N_{\rm H}}$ & $kT_1$ & $kT_2$ & $\log{EM_1}$       & $\log{EM_2}$       & $\log{L_{\rm x}}^*$ \\
   &        &                          & [${\rm cm^{-2}}$] & [keV]  & [keV]  & [${\rm cm}^{-3}$]  & [${\rm cm}^{-3}$]  & [erg/s]             \\ \hline
HD\,32964  & B &   $0.87$ ( $ 5$) & $-$     &  $ 0.54 $ &  $ 1.62 $ & $ 52.24 $ & $ 52.16 $ & $  29.41 $ \\
HD\,73952  & B &   $1.13$ ( $12$) & $-$     &  $ 0.68 $ &  $ 1.26 $ & $ 52.19 $ & $ 51.97 $ & $  29.33 $ \\
HD\,73952  & A &   $0.87$ ( $ 4$) & $-$     &  $ 0.32 $ &  $ 1.04 $ & $ 51.98 $ & $ 51.75 $ & $  29.04 $ \\
HD\,110073 & B &   $0.32$ ( $ 4$) & $-$     &  $ 0.62 $ &  $ 1.03 $ & $ 52.36 $ & $ 51.79 $ & $  29.35 $ \\
HD\,134946 & A &   $0.80$ ( $10$) & $-$     &  $ 0.69 $ &  $ 1.30 $ & $ 52.68 $ & $ 52.63 $ & $  29.91 $ \\
HD\,129791 & A &   $0.89$ ( $47$) & $21.23$ &  $ 0.64 $ &  $ 1.49 $ & $ 53.38 $ & $ 53.02 $ & $  30.36 $ \\
HD\,129791 & L &   $0.63$ ( $13$) & $-$     &  $ 0.51 $ &  $ 1.15 $ & $ 52.42 $ & $ 52.19 $ & $  29.53 $ \\
\hline
HD\,113703 & L &   $1.15$ ( $73$) & $-$     &  $ 0.80 $ &  $ 4.32 $ & $ 53.15 $ & $ 52.87 $ & $  30.30 $ \\
HD\,113703 & B &   $1.83$ ( $40$) & $-$     &  $ 0.34 $ &  $ 0.99 $ & $ 52.83 $ & $ 52.39 $ & $  29.77 $ \\
HD\,1685   & B &   $0.95$ ( $ 8$) & $-$     &  $ 0.70 $ & $>0.88  $ & $ 51.89 $ & $ 52.01 $ & $  29.40 $ \\
HD\,1685   & A &   $1.22$ ( $ 3$) & $-$     &  $ 0.25 $ &  $ 1.06 $ & $ 52.34 $ & $ 51.69 $ & $  29.12 $ \\
HD\,123445 & B/C & $0.79$ ( $ 6$) & $-$     &  $ 0.43 $ &  $ 0.98 $ & $ 52.20 $ & $ 52.75 $ & $  29.94 $ \\
HD\,133880 & B &   $1.95$ ( $13$) & $-$     &  $ 0.20 $ &  $ 1.00 $ & $ 52.79 $ & $ 52.72 $ & $  29.94 $ \\
HD\,169978 & A &   $0.87$ ( $ 7$) & $-$     &  $ 0.38 $ &  $ 1.29 $ & $ 53.01 $ & $ 52.61 $ & $  29.95 $ \\
\hline
\multicolumn{9}{l}{$^*$ in the $0.5-8$\,keV passband; $L_{\rm x}$ refers to the distance given in Table~\ref{tab:obslog}.} \\
\end{tabular}
\end{center}\end{table*}